\documentclass[usegraphicx]{mn2e}

\title[Does $Q$ correlate with $j$ or $M$\,?]
      {Angular momentum and local gravitational instability in galaxy discs:
       does $Q$ correlate with $j$ or $M$\,?}

\author[A. B. Romeo and K. M. Mogotsi]
       {Alessandro B. Romeo$^{1}$\thanks{E-mail: romeo@chalmers.se}
        and Keoikantse Moses Mogotsi$^{2}$\\
        $^{1}$Department of Space, Earth and Environment,
              Chalmers University of Technology,
              SE-41296 Gothenburg, Sweden\\
        $^{2}$South African Astronomical Observatory,
              PO Box 9, Observatory, Cape Town, 7935 South Africa}

\begin{document}

\date{Accepted 2018 June 26.
      Received 2018 June 25; in original form 2018 May 15}

\pagerange{\pageref{firstpage}--\pageref{lastpage}}

\pubyear{2018}

\maketitle

\label{firstpage}

\begin{abstract}
We introduce a new diagnostic for exploring the link between angular momentum
and local gravitational instability in galaxy discs.  Our diagnostic
incorporates the latest developments in disc instability research, is fully
consistent with approximations that are widely used for measuring the stellar
specific angular momentum, $j_{\star}=J_{\star}/M_{\star}$, and is also very
simple.  We show that such a disc instability diagnostic hardly correlates
with $j_{\star}$ or $M_{\star}$, and is remarkably constant across spiral
galaxies of any given type (Sa--Sd), stellar mass
($M_{\star}=10^{9.5}\mbox{--}10^{11.5}\,\mbox{M}_{\odot}$) and velocity
dispersion anisotropy ($\sigma_{z\star}/\sigma_{R\star}=0\mbox{--}1$).  The
fact that $M_{\star}$ is tightly correlated with star formation rate
($\mathrm{SFR}$), molecular gas mass ($M_{\mathrm{mol}}$), metallicity
($12+\log\mathrm{O/H}$) and other fundamental galaxy properties thus implies
that nearby star-forming spirals self-regulate to a quasi-universal disc
stability level.  This proves the existence of the self-regulation process
postulated by several star formation models, but also raises important
caveats.
\end{abstract}

\begin{keywords}
instabilities --
stars: kinematics and dynamics --
ISM: kinematics and dynamics --
galaxies: ISM --
galaxies: kinematics and dynamics --
galaxies: star formation.
\end{keywords}

\section{INTRODUCTION}

Today, 35 years after the pioneering work of Fall (1983), angular momentum is
regarded as one of the most fundamental galaxy properties.  Fall's scaling
law $j_{\star}\propto M_{\star}^{2/3}$, which links the stellar specific
angular momentum ($j_{\star}=J_{\star}/M_{\star}$) to the stellar mass
($M_{\star}$), has been confirmed and refined in a wide variety of contexts,
and forms the basis of a new physical-morphological classification of
galaxies (e.g., Romanowsky \& Fall 2012; Obreschkow \& Glazebrook 2014; Elson
2017; Lagos et al.\ 2017; Lapi et al.\ 2018; Posti et al.\ 2018; Sweet et
al.\ 2018).  Angular momentum is linked to global dynamical processes such as
the formation and evolution of galaxies, and the gravitational instability of
galaxy discs to bar formation (e.g., Mo et al.\ 1998; Athanassoula 2008;
Agertz \& Kravtsov 2016; Sellwood 2016; Okamura et al.\ 2018; Zoldan et
al.\ 2018).

There is recent evidence that angular momentum is also linked to local disc
instability.  Using the \textsc{Dark Sage} semi-analytic model of galaxy
evolution, Stevens et al.\ (2016) showed that disc instabilities are crucial
for regulating both the mass and the spin of galaxy discs.  Obreschkow et
al.\ (2016) found that the mass fraction of neutral atomic gas in isolated
local disc galaxies can be described by a hybrid stability model, which
combines the H\,\textsc{i} velocity dispersion with the mass and specific
angular momentum of the whole (gas+stars) disc.  Such a stability model was
used by Lutz et al.\ (2018) to analyse galaxies that are extremely rich in
H\,\textsc{i}, and to associate their high H\,\textsc{i} content with their
high specific angular momentum.  Zasov \& Zaitseva (2017) showed that the
relation between atomic gas mass and disc specific angular momentum in
late-type star-forming galaxies is equally well described by a simpler
stability model controlled by the gas Toomre parameter.  Zasov \& Zaitseva
(2017) also discussed the impact that radial variation in the gas velocity
dispersion may have on their model, and the role that stars may play in that
scenario.  Swinbank et al.\ (2017) found that angular momentum plays a major
role in defining the stability of galaxy discs at $z\sim1$, and identified a
correlation between the stellar specific angular momentum and the gas Toomre
parameter.  Other pieces of evidence are discussed by Lagos et al.\ (2017)
and Swinbank et al.\ (2017).

In spite of such evidence, there is still no tight constraint on the link
between angular momentum and local gravitational instability in galaxy discs.
Note, in fact, that diagnostics like the gas Toomre parameter are highly
unreliable indicators of gravitational instability.  This concerns not only
nearby spirals, where disc instabilities are driven by stars (Romeo \&
Mogotsi 2017), but also gas-rich galaxies at low and high redshifts, where
turbulence can drive the disc into regimes that are far from Toomre/Jeans
instability (Romeo et al.\ 2010; Romeo \& Agertz 2014).

This Letter provides the astronomical community with a simple and reliable
diagnostic for exploring this link in nearby spirals.  Besides deriving such
a diagnostic and comparing it with other stability parameters (Sect.\ 2), we
illustrate its strength with an eloquent example, which tightly constrains
the relation between disc stability level, stellar specific angular momentum
and stellar mass (Sect.\ 3).  This turns out to have wider implications,
which we discuss together with our conclusions (Sect.\ 4).

\section{DISC INSTABILITY DIAGNOSTIC}

\subsection{The route to $\langle\mathcal{Q}_{\star}\rangle$}

To explore the link between angular momentum and local gravitational
instability in nearby star-forming spirals, we need a reliable disc
instability diagnostic.  Contrary to what is commonly assumed, the gas Toomre
parameter is not a reliable diagnostic: stars, and not molecular or atomic
gas, are the primary driver of disc instabilities in spiral galaxies, at
least at the spatial resolution of current extragalactic surveys (Romeo \&
Mogotsi 2017).  This is confirmed by other investigations (Marchuk 2018;
Marchuk \& Sotnikova 2018; Mogotsi \& Romeo 2018), and is true even for a
powerful starburst+Seyfert galaxy like NGC 1068 (Romeo \& Fathi 2016).  The
stellar Toomre parameter is a more reliable diagnostic, but it does not
include the stabilizing effect of disc thickness, which is important and
should be taken into account (Romeo \& Falstad 2013).  The simplest
diagnostic that does this accurately is the Romeo-Falstad $\mathcal{Q}_{N}$
stability parameter for one-component ($N=1$) stellar ($\star$) discs, which
we consider as a function of galactocentric distance $R$:
\begin{equation}
\mathcal{Q}_{\star}(R)=
Q_{\star}(R)\,T_{\star}\,,
\end{equation}
where $Q_{\star}=\kappa\sigma_{\star}/\pi G\Sigma_{\star}$ is the stellar
Toomre parameter ($\sigma$ denotes the radial velocity dispersion), and
$T_{\star}$ is a factor that encapsulates the stabilizing effect of disc
thickness for the whole range of velocity dispersion anisotropy
($\sigma_{z}/\sigma_{R}$) observed in galactic discs:
\begin{equation}
T_{\star}=
\left\{\begin{array}{ll}
       {\displaystyle1+0.6\left(\frac{\sigma_{z}}{\sigma_{R}}\right)_{\star}^{2}}
                       &
       \mbox{if\ }0\leq(\sigma_{z}/\sigma_{R})_{\star}\leq0.5\,,
                         \\
                       & \\
       {\displaystyle0.8+0.7\left(\frac{\sigma_{z}}{\sigma_{R}}\right)_{\star}}
                       &
       \mbox{if\ }0.5\leq(\sigma_{z}/\sigma_{R})_{\star}\leq1\,.
       \end{array}
\right.
\end{equation}
Observations do not yet constrain the radial variation of
$(\sigma_{z}/\sigma_{R})_{\star}$, hence that of $T_{\star}$ (Gerssen \&
Shapiro Griffin 2012; Marchuk \& Sotnikova 2017; Pinna et al.\ 2018).

As $\mathcal{Q}_{\star}(R)$ is a local quantity, it cannot be directly
related to the stellar specific angular momentum,
\begin{equation}
j_{\star}=
\frac{1}{M_{\star}}
\int_{0}^{\infty}Rv_{\mathrm{c}}(R)\,\Sigma_{\star}(R)\,2\pi R\,\mathrm{d}R
\end{equation}
(e.g., Romanowsky \& Fall 2012).  This equation tells us that $j_{\star}$ is
the mass-weighted average of $Rv_{\mathrm{c}}(R)$, the orbital specific
angular momentum.  So it is natural to consider the mass-weighted average of
$\mathcal{Q}_{\star}(R)$.  Current integral-field-unit (IFU) surveys allow
deriving reliable radial profiles of $\mathcal{Q}_{\star}$ up to $R\approx
R_{\mathrm{e}}$, the effective (half-light) radius.  This limit is imposed by
the sparsity of reliable $\sigma_{\star}$ measurements for $R\ga
R_{\mathrm{e}}$ (Martinsson et al.\ 2013; Falc\'{o}n-Barroso et al.\ 2017;
Mogotsi \& Romeo 2018).  In view of these facts, we take the mass-weighted
average of $\mathcal{Q}_{\star}(R)$ over one effective radius:
\begin{equation}
\langle\mathcal{Q}_{\star}\rangle=
\frac{1}{M_{\star}(R_{\mathrm{e}})}
\int_{0}^{R_{\mathrm{e}}}\mathcal{Q}_{\star}(R)\,\Sigma_{\star}(R)\,2\pi R\,\mathrm{d}R\,.
\end{equation}
This ensures that $\langle\mathcal{Q}_{\star}\rangle$ and $j_{\star}$ have a
similar relation to their local counterparts, which simplifies the following
analysis.

To illustrate the usefulness of Eq.\ (4), let us calculate
$\langle\mathcal{Q}_{\star}\rangle$ for a galaxy model that is behind the
simple, accurate and widely used approximation
$j_{\star}=1.19\,R_{\mathrm{e}}v_{\mathrm{c}}$: an exponential disc having a
constant mass-to-light ratio and rotating at a constant circular speed (e.g.,
Romanowsky \& Fall 2012).  For this galaxy model,
$M_{\star}(R_{\mathrm{e}})=\frac{1}{2}M_{\star}$ and
$\kappa(R)=\sqrt{2}\,v_{\mathrm{c}}/R$ (see, e.g., Binney \& Tremaine 2008),
which can be expressed in terms of $j_{\star}$ using the approximation above.
The resulting $\langle\mathcal{Q}_{\star}\rangle$ is given by
\begin{equation}
\langle\mathcal{Q}_{\star}\rangle=
4.75\,\frac{j_{\star}\overline{\sigma}_{\star}}{GM_{\star}}\,\,T_{\star}\,,
\end{equation}
where $j_{\star}$ is the total stellar specific angular momentum and
$M_{\star}$ is the total stellar mass, while $\overline{\sigma}_{\star}$ is
the radial average of $\sigma_{\star}(R)$ over one effective radius:
\begin{equation}
\overline{\sigma}_{\star}=
\frac{1}{R_{\mathrm{e}}}
\int_{0}^{R_{\mathrm{e}}}\sigma_{\star}(R)\,\mathrm{d}R\,.
\end{equation}
Varying the radius over which $\mathcal{Q}_{\star}(R)$ and
$\sigma_{\star}(R)$ are averaged has a remarkably weak effect on the
numerical factor in Eq.\ (5): if one averages over $2R_{\mathrm{e}}$ (rather
than $R_{\mathrm{e}}$), then the numerical factor is 5.60 (rather than 4.75).
Averaging over $2R_{\mathrm{e}}$ requires reliable $\sigma_{\star}$
measurements up to such radii, which are currently very sparse (Martinsson et
al.\ 2013; Falc\'{o}n-Barroso et al.\ 2017; Mogotsi \& Romeo 2018) but will
proliferate with the advent of second-generation IFU surveys using the Multi
Unit Spectroscopic Explorer (MUSE).  This is different from the case of
$j_{\star}$ measurements, which have already entered the high-precision era
(e.g., Obreschkow \& Glazebrook 2014; Lapi et al.\ 2018; Posti et al.\ 2018).

\subsection{$\langle\mathcal{Q}_{\star}\rangle$
            versus other stability parameters}

$\langle\mathcal{Q}_{\star}\rangle$ measures the local stability of galaxy
discs in an averaged, mass-weighted sense.  Since
$\langle\mathcal{Q}_{\star}\rangle$ depends on mass and specific angular
momentum, and since these quantities also affect the stability of galaxy
discs against bar formation (Mo et al.\ 1998),
$\langle\mathcal{Q}_{\star}\rangle$ must be related to the
Efstathiou-Lake-Negroponte global stability parameter,
\begin{equation}
\epsilon_{\mathrm{m}}\equiv
\frac{V_{\mathrm{max}}}{(GM_{\mathrm{d}}/R_{\mathrm{d}})^{1/2}}\,,
\end{equation}
where $V_{\mathrm{max}}$ is the maximum rotation velocity, $M_{\mathrm{d}}$
is the mass of the disc, and $R_{\mathrm{d}}$ is the disc scale length
(Efstathiou et al.\ 1982; Christodoulou et al.\ 1995).  For the galaxy model
that leads to Eq.\ (5), we get: $V_{\mathrm{max}}=v_{\mathrm{c}}$,
$M_{\mathrm{d}}=M_{\star}$ and $R_{\mathrm{d}}=j_{\star}/2v_{\mathrm{c}}$
(e.g., Romanowsky \& Fall 2012), hence
\begin{equation}
\langle\mathcal{Q}_{\star}\rangle\approx
\epsilon_{\mathrm{m}}^{2}\;
(10\,\overline{\sigma}_{\star}/v_{\mathrm{c}})\;T_{\star}\,.
\end{equation}
In other words, $\langle\mathcal{Q}_{\star}\rangle$ can be viewed as
$\epsilon_{\mathrm{m}}^{2}$ altered by two factors: the first one,
$\approx(10\,\overline{\sigma}_{\star}/v_{\mathrm{c}})$, results from the
different roles that random and ordered motions play in local and global
gravitational instabilities; the second one, $T_{\star}$, represents the
stabilizing effect of disc thickness, which depends on the velocity
dispersion anisotropy (see Eq.\ 2).

$\langle\mathcal{Q}_{\star}\rangle$ is not the only parameter that relates
local disc stability to mass and specific angular momentum.  The first
attempt to do that was made by Obreschkow \& Glazebrook (2014).  Using
dimensional analysis and physical insight, they defined a disc-averaged
Toomre parameter as $\overline{Q}\propto\sigma_{0}jM^{-1}$, where
$\sigma_{0}$ is a velocity dispersion scale.  Obreschkow et al.\ (2016)
redefined $\overline{Q}$ as $q\equiv
j_{\mathrm{disc}}\,\sigma_{\mathrm{HI}}/(GM_{\mathrm{disc}})$ and referred to
this hybrid quantity as a `global' disc stability parameter.%
\footnote{What Obreschkow et al.\ (2016) actually meant by `global' was
  `mass-weighted average'.  In fact, $q$ does not concern global disc
  stability against bar or spiral structure formation.}
The stability criterion also changed from $\overline{Q}\geq1$ (Obreschkow et
al.\ 2015) to $q\ga1/(\sqrt{2}\,\mathrm{e})$ or $q\ga0.4$ (Obreschkow et
al.\ 2016), depending on the model.  Although
$\langle\mathcal{Q}_{\star}\rangle$ may look similar to $\overline{Q}$ and
$q$, it is not.  First of all, $\langle\mathcal{Q}_{\star}\rangle$ is a
robustly defined parameter, which results from state-of-the-art diagnostics
for detecting gravitational instabilities in galaxy discs (see Sect.\ 2.1).
Second, $\langle\mathcal{Q}_{\star}\rangle$ depends on
$\overline{\sigma}_{\star}$, which differs radically from
$\sigma_{\mathrm{HI}}$ not only in value but also in meaning: disc
instabilities in spiral galaxies are driven by stars, not by atomic gas (see
again Sect.\ 2.1).

\section{PRACTICAL USE OF $\langle\mathcal{Q}_{\star}\rangle$}

\subsection{Exploring the
            $\langle\mathcal{Q}_{\star}\rangle$--$M_{\star}$--$j_{\star}$
            correlation}

Now that we have a reliable disc instability diagnostic, let us explore how
$\langle\mathcal{Q}_{\star}\rangle$ correlates with $M_{\star}$ and
$j_{\star}$.  To do this, we make use of Eq.\ (5) and the following scaling
relations:
\begin{itemize}
\item $\log j_{\star}=0.52\,(\log M_{\star}-11)+3.18$, which has an rms
  scatter of 0.19 dex (Romanowsky \& Fall 2012);
\item $\log\overline{\sigma}_{\star}=0.45\log M_{\star}-2.77$, which has an
  rms scatter of 0.10 dex (Mogotsi \& Romeo 2018).
\end{itemize}
These scaling relations are least-squares fits to accurate measurements of
$j_{\star}\;[\mbox{kpc}\;\mbox{km\,s}^{-1}]$ and
$\overline{\sigma}_{\star}\;[\mbox{km\,s}^{-1}]$ versus
$M_{\star}\;[\mbox{M}_{\odot}]$, and are applicable in tandem to spiral
galaxies of type Sa--Sd and stellar mass
$M_{\star}\approx10^{9.5}\mbox{--}10^{11.5}\,\mbox{M}_{\odot}$.  Contrary to
$j_{\star}$ and $\overline{\sigma}_{\star}$, $T_{\star}$ is uncorrelated with
$M_{\star}$.  This follows from the facts that
$(\sigma_{z}/\sigma_{R})_{\star}$ is uncorrelated with Hubble type (Pinna et
al.\ 2018) and Hubble type is strongly correlated with $M_{\star}$ (e.g.,
Conselice 2006).  If we regard the $j_{\star}$--$M_{\star}$ and
$\overline{\sigma}_{\star}$--$M_{\star}$ best-fitting relations as functional
relations and the associated rms scatters as uncorrelated, then the expected
$\langle\mathcal{Q}_{\star}\rangle$--$M_{\star}$ scaling relation is
\begin{equation}
\langle\mathcal{Q}_{\star}\rangle=
5.4\left(\frac{M_{\star}}{\mbox{M}_{\odot}}\right)^{-0.03}T_{\star}
\end{equation}
and has an rms scatter of approximately 0.21 dex
$(0.21=\sqrt{0.19^{2}+0.10^{2}})$, i.e.\ an rms scatter of approximately a
factor of 1.6.  Inverting the $j_{\star}$--$M_{\star}$ relation, we can also
infer $\langle\mathcal{Q}_{\star}\rangle$ as a function of $j_{\star}$:
\begin{equation}
\langle\mathcal{Q}_{\star}\rangle=
3.9\left(\frac{j_{\star}}{1\;\mbox{kpc}\;\mbox{km\,s}^{-1}}\right)^{-0.06}T_{\star}\,.
\end{equation}
Hereafter we will focus on Eq.\ (9), since $M_{\star}$ is a more classical
observable than $j_{\star}$.%
\footnote{Obreschkow et al.\ (2016) found that $q\propto
  M_{\mathrm{disc}}^{-1/3}$, but this cannot be compared with our
  $\langle\mathcal{Q}_{\star}\rangle$--$M_{\star}$ scaling relation since $q$
  and $\langle\mathcal{Q}_{\star}\rangle$ are conceptually different
  parameters (see Sect.\ 2.2).}

Eq.\ (9) predicts that a two-orders-of-magnitude variation in $M_{\star}$, as
observed across spiral galaxies of type Sa--Sd, `collapses' into a $<20\%$
variation in $\langle\mathcal{Q}_{\star}\rangle$:
\begin{equation}
M_{\star}=10^{9.5}\mbox{--}10^{11.5}\,\mbox{M}_{\odot}
\;\;\;\Longrightarrow\;\;\;
\langle\mathcal{Q}_{\star}\rangle\simeq
2.4\mbox{--}2.8\,\,T_{\star}\,.
\end{equation}
The observed variation in $(\sigma_{z}/\sigma_{R})_{\star}$ has a more
significant impact, but the total expected variation in
$\langle\mathcal{Q}_{\star}\rangle$ is still within a factor of two:
\begin{equation}
(\sigma_{z}/\sigma_{R})_{\star}=0\mbox{--}1
\;\;\;\Longrightarrow\;\;\;
\langle\mathcal{Q}_{\star}\rangle\sim
2\mbox{--}4\,.
\end{equation}
The prediction that $\langle\mathcal{Q}_{\star}\rangle$ has an expected value
of $\sim2\mbox{--}4$ for spiral galaxies of any given type, stellar mass and
velocity dispersion anisotropy is in remarkable agreement with high-quality
measurements of the disc stability level in such galaxies (e.g., Westfall et
al.\ 2014; Hallenbeck et al.\ 2016; Garg \& Banerjee 2017; Romeo \& Mogotsi
2017; Marchuk 2018; Marchuk \& Sotnikova 2018).  An expected value of
$\langle\mathcal{Q}_{\star}\rangle\sim2\mbox{--}4$ is also meaningful from a
theoretical point of view: it tells us that spiral galaxies are, in a
statistical sense, marginally stable against non-axisymmetric perturbations
(e.g., Griv \& Gedalin 2012) and gas dissipation (Elmegreen 2011), although
the precise value of the critical stability level is still questioned (Romeo
\& Fathi 2015).

\subsection{Non-correlation confirmed}

To test the robustness of our results, we analyse a sample of 34 nearby
spiral galaxies of type Sa--Sd from the Calar Alto Legacy Integral Field Area
(CALIFA) survey, as listed in table 1 of Mogotsi \& Romeo (2018).  These are
galaxies with accurate measurements of the epicyclic frequency $\kappa$
(Kalinova et al.\ 2017; Mogotsi \& Romeo 2018), stellar radial velocity
dispersion $\sigma_{\star}$ (Falc\'{o}n-Barroso et al.\ 2017; Mogotsi \&
Romeo 2018), stellar velocity dispersion anisotropy
$\sigma_{z\star}/\sigma_{R\star}$ (Kalinova et al.\ 2017), stellar mass
$M_{\star}$ and other galaxy properties (Bolatto et al.\ 2017).  These are
all the quantities needed to compute $\langle\mathcal{Q}_{\star}\rangle$ from
Eq.\ (4), except for the stellar mass surface density $\Sigma_{\star}$, which
has not been measured in many galaxies of our sample (S\'{a}nchez et
al.\ 2016).  Note, however, that $\Sigma_{\star}$ only enters the
normalization factor $M_{\star}(R_{\mathrm{e}})$ in Eq.\ (4), which is close
to $\frac{1}{2}M_{\star}$ (Gonz\'{a}lez Delgado et al.\ 2014).  So we use
this approximation, but compute the integral in Eq.\ (4) accurately by taking
into account the Voronoi binning of CALIFA data (see Cappellari 2009 for a
review).  In simple words, we sum over Voronoi bins rather than over circular
rings.

%%%%%%%%%%%%%%%%%%%%%%%%%%%%%%%%%%%%%%%%%%%%%%%%%%%%%%%%%%%%%%%%%%%%%%%%%%%%%
\begin{figure}
\includegraphics[scale=1.]{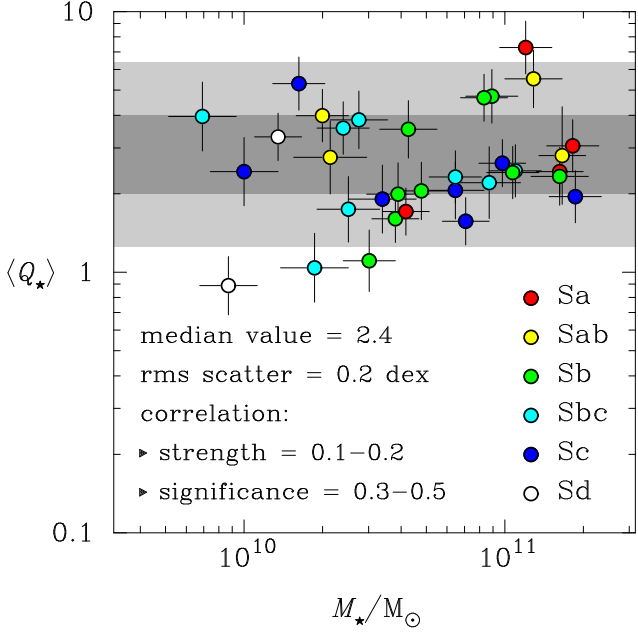}
\caption{Our disc instability diagnostic,
  $\langle\mathcal{Q}_{\star}\rangle$, versus stellar mass, $M_{\star}$, for
  a sample of 34 nearby spiral galaxies of type Sa--Sd (colour-coded) from
  the CALIFA survey.  The dark grey area shows the variation in
  $\langle\mathcal{Q}_{\star}\rangle$ predicted by Eq.\ (9), while the light
  grey area shows the rms scatter around this range of values predicted in
  Sect.\ 3.1.  Statistical information about the data is given in summary
  form and simplified notation (see Sect.\ 3.2 for more information).}
\end{figure}
%%%%%%%%%%%%%%%%%%%%%%%%%%%%%%%%%%%%%%%%%%%%%%%%%%%%%%%%%%%%%%%%%%%%%%%%%%%%%

Fig.\ 1 illustrates that the resulting $\langle\mathcal{Q}_{\star}\rangle$
versus $M_{\star}$ is fully consistent with the predictions made in
Sect.\ 3.1.  $\langle\mathcal{Q}_{\star}\rangle$ has a median value of 2.4,
which is within the expected range of values ($\sim2\mbox{--}4$), and has an
rms scatter of approximately 0.2 dex, which is close to the expected one
(0.21 dex).  Fig.\ 1 also shows that there is no clear correlation between
$\langle\mathcal{Q}_{\star}\rangle$ and $M_{\star}$.  To quantify the
strength and the significance of a possible
$\langle\mathcal{Q}_{\star}\rangle$--$M_{\star}$ correlation, we present the
results of three statistical measures and associated tests (see, e.g., Press
et al.\ 1992).  We find that:
\begin{enumerate}
\item Pearson's correlation coefficient $r=0.17$, and its significance level
  $p_{r}=0.32$;
\item Spearman's rank correlation coefficient $\rho=0.11$, and its two-sided
  significance level $p_{\rho}=0.53$;
\item Kendall's rank correlation coefficient $\tau=0.12$, and its two-sided
  significance level $p_{\tau}=0.31$.
\end{enumerate}
These numbers speak clearly: $\langle\mathcal{Q}_{\star}\rangle$ hardly
correlates with $M_{\star}$, as predicted in Sect.\ 3.1.

\section{CONCLUSIONS}

\begin{itemize}
\item If there is a direct link between angular momentum and local
  gravitational instability in nearby star-forming spirals, then it must
  involve $j_{\star}$ and $\mathcal{Q}_{\star}=Q_{\star}T_{\star}$.  This is
  because stars ($\star$), and not molecular or atomic gas, play the leading
  role in the disc instability scenario, and because disc thickness has an
  important stabilizing effect ($T_{\star}$).
\item Since $j_{\star}$ is the mass-weighted average of a local quantity,
  $Rv_{\mathrm{c}}(R)$, and since $\mathcal{Q}_{\star}$ itself is a local
  quantity, an unbiased relation must involve $j_{\star}$ and
  $\langle\mathcal{Q}_{\star}\rangle$, the mass-weighted average of
  $\mathcal{Q}_{\star}$.
\item This Letter introduces a new disc instability diagnostic that satisfies
  the two requirements above, and which is simple and fully consistent with
  the widely used approximation $j_{\star}=2R_{\mathrm{d}}V$ (see Eq.\ 5).
  Although conceptually distinct, our diagnostic is related to the
  Efstathiou-Lake-Negroponte global stability parameter via the degree of
  rotational support, $V/\sigma$, and the velocity dispersion anisotropy,
  $\sigma_{z}/\sigma_{R}$ (see Eq.\ 8).
\item Making use of previously established scaling relations, we show that
  $\langle\mathcal{Q}_{\star}\rangle$ hardly correlates with $j_{\star}$ or
  $M_{\star}$: $\langle\mathcal{Q}_{\star}\rangle\propto
  j_{\star}^{-0.06}\propto M_{\star}^{-0.03}$ (see Eqs 9 and 10).  This
  scaling relation results in a remarkably constant
  $\langle\mathcal{Q}_{\star}\rangle\sim2\mbox{--}4$ across spiral galaxies
  of any given type (Sa--Sd), stellar mass
  ($M_{\star}=10^{9.5}\mbox{--}10^{11.5}\,\mbox{M}_{\odot}$) and velocity
  dispersion anisotropy ($\sigma_{z\star}/\sigma_{R\star}=0\mbox{--}1$).
  These results are fully consistent with high-quality measurements of the
  disc stability level in such galaxies, and with theoretical estimates of
  the local stability threshold in galaxy discs.  The robustness of our
  results is further confirmed by a detailed analysis of a sample of 34
  nearby spirals from the CALIFA survey.  Details are given in Sect.\ 3.
\end{itemize}

Our results have wider implications.  It is well known that $M_{\star}$ is
tightly correlated with star formation rate ($\mathrm{SFR}$), molecular gas
mass ($M_{\mathrm{mol}}$), metallicity ($12+\log\mathrm{O/H}$) and other
fundamental galaxy properties (e.g., Conselice 2006; Nagamine et al.\ 2016;
Lapi et al.\ 2018).  The fact that $\langle\mathcal{Q}_{\star}\rangle$ varies
very weakly with $M_{\star}$ thus implies that nearby star-forming spirals
self-regulate to a quasi-universal disc stability level.  This is
conceptually similar to the self-regulation process postulated by several
star formation models, which assume $Q=1$ throughout the disc (see sect.\ 1
of Krumholz et al.\ 2018 for an overview).  Note, however, that there are two
significant differences.  First of all, the key quantity is basically
$\mathcal{Q}_{\star}$ and not the gas Toomre parameter
$Q_{\mathrm{g}}=\kappa\sigma_{\mathrm{g}}/\pi G\Sigma_{\mathrm{g}}$.  In
fact, $Q_{\mathrm{g}}$ varies by more than one order of magnitude in nearby
star-forming spirals (see fig.\ 5 of Romeo \& Wiegert 2011).  Second,
$\mathcal{Q}_{\star}$ is well above unity and is approximately constant
($\sim2\mbox{--}4$) only in a statistical sense.  In fact,
$\mathcal{Q}_{\star}$ can vary by more than a factor of two even within an
individual spiral galaxy (see fig.\ A.14 of Grebovi\'{c} 2014).
New-generation star formation models must take these two facts into account,
and a significant step forward has just been taken (Krumholz et al.\ 2018).

Finally, the practical use of $\langle\mathcal{Q}_{\star}\rangle$ extends
beyond the eloquent example illustrated in this Letter.  Since angular
momentum and local gravitational instability are key ingredients in the
formation and evolution of galaxy discs (e.g., Lagos et al.\ 2017; Krumholz
et al.\ 2018), $\langle\mathcal{Q}_{\star}\rangle$ can indeed be used in a
variety of contexts.  One such application could be to constrain the relation
between angular momentum, galaxy morphology and star formation more tightly
than now, which is a primary goal in galactic angular momentum research
(e.g., Obreschkow \& Glazebrook 2014; Obreschkow et al.\ 2015; Lagos et
al.\ 2017; Swinbank et al.\ 2017).  This requires reliable measurements of
the disc stability level, which $\langle\mathcal{Q}_{\star}\rangle$ has been
shown to provide.

\section*{ACKNOWLEDGEMENTS}

ABR dedicates this Letter to his mother Grazia: in your memory, with infinite
love and sorrow.  We are very grateful to Oscar Agertz, Claudia Lagos, Robert
Nau, Lorenzo Posti, Florent Renaud and Anatoly Zasov for useful discussions.
We are also grateful to an anonymous referee for insightful comments and
suggestions, and for encouraging future work on the topic.  This work made
use of data from the CALIFA survey (http://califa.caha.es/).

\bsp

\label{lastpage}

\end{document}